\RequirePackage{fix-cm}

\documentclass[twocolumn]{svjour3}         
\usepackage{braket}
\usepackage{ulem}
\usepackage{xcolor}
\expandafter\let\csname equation*\endcsname=\relax 
\expandafter\let\csname endequation*\endcsname=\relax
\usepackage{amsmath,amssymb,float}
\usepackage{subfig}
\usepackage{graphicx}
\usepackage[colorlinks=true,linkcolor={blue},citecolor={red},
            breaklinks=true
            ]{hyperref}

\smartqed   

\begin{document}

\title{Information processing capacity of spin-based quantum reservoir computing systems}

\author{R. Mart\'inez-Pe\~na, J. Nokkala, G. L. Giorgi, R. Zambrini, M. C. Soriano }

\institute{Authors are with the Instituto de F\'{i}sica Interdisciplinar y Sistemas Complejos (IFISC, UIB-CSIC), Campus Universitat de les Illes Balears E-07122, Palma de Mallorca, Spain.\\
Corresponding author: R. Mart\'inez-Pe\~na (\email{rmartinez@ifisc.uib-csic.es})
}

\maketitle

\begin{abstract}
The  dynamical behaviour of complex quantum systems can be harnessed for information processing.  With this aim, quantum reservoir computing (QRC) with Ising spin networks was recently introduced as a quantum version of classical reservoir computing.
In turn, reservoir computing is a neuro-inspired machine learning technique that consists in exploiting dynamical systems to solve nonlinear and temporal tasks.
We characterize the performance of the spin-based QRC model with the Information Processing Capacity (IPC), which allows to quantify the computational capabilities of a dynamical system beyond specific tasks.
The influence on the IPC of the input injection frequency, time multiplexing, and different measured observables encompassing local spin measurements as well as correlations, is addressed. 
We find conditions for an optimum input driving and provide different alternatives for the choice of the output variables used for the readout.
This  work establishes a clear picture of the computational capabilities of a quantum network of spins for reservoir computing. 
Our results pave the way to future research on QRC both from the theoretical and experimental points of view.


\end{abstract}

\section{Introduction}

Machine learning has become one of the fastest-growing research lines in the last years, with deep learning being a prominent example \cite{goodfellow2016deep}. Applied to many fields like computer vision \cite{krizhevsky2012imagenet}, physical sciences \cite{carleo2019machine}, medicine \cite{hinton2018deep} or language processing \cite{young2018recent}, machine learning techniques enable us to solve problems that were very hard or even impossible to tackle with more traditional tools. A specific group of problems that belongs to this category are the ones that involve the processing of temporal signals, such as speech recognition \cite{triefenbach2014large}, time series prediction \cite{pathak2018model} or channel equalization \cite{antonik2016online}.

The general concept of Recurrent Neural Network (RNN) encloses one of the main techniques employed nowadays to solve temporal tasks \cite{makridakis2018m4}. A RNN is a neural network whose nodes are recursively connected, which allows information to remain in the network through time, giving the system ``memory". Among all the RNN techniques, Reservoir Computing (RC)  is a promising line of research that exploits dynamical systems to process the input information \cite{maass2002real,lukovsevivcius2009reservoir}.  

The inception of this field comes from two different approaches with different motivations. On the one hand, Echo State Networks \cite{jaeger2001echo} were designed with the aim of reducing the complexity of the training for RNNs while keeping a good performance. On the other hand,
Liquid State Machines  were proposed almost at the same time with the focus on modeling neural microcircuits \cite{maass2002real}. 
Both of them share the characteristic of not relying on the tuning of the parameters of the recurrent network, also known as reservoir in this context. Thus, training during the learning process only adapts the weights of the output layer, where the state of the reservoir is read out. 

By training only the output weights, the reservoir computing paradigm requires certain conditions for a proper performance. These conditions
are usually referred to as the Echo State Property (ESP), the Fading Memory Property (FMP) and the Separability Property (SP) \cite{verstraeten2007experimental}. The state of a driven dynamical system with the ESP is uniquely determined by the input history. FMP implies that the dynamical system dissipates the information about the input in time. Both ESP and FMP are deeply connected and they require to forget the initial state of the reservoir as a necessary condition. Finally, the SP enforces that the reservoir computer produces different outputs for any pair of different inputs.  

The simplifications in the training algorithm also bring important benefits, in addition to a direct speed-up in the training time. First, it opens the possibility of solving many tasks with the same reservoir, training only the output layer independently for each task. Second and of practical relevance, since no fine tuning is required for the reservoir, a large amount of possible implementations for information processing have been demonstrated in numerical simulations and even using physical hardware \cite{tanaka2019recent}. 

To name a few, the list of experimental realizations include photonic \cite{van2017advances}, electronic \cite{appeltant2011information} and spintronic \cite{torrejon2017neuromorphic}  implementations. However, quantum implementations are visibly lacking. That is why Quantum Reservoir Computing (QRC) was introduced recently  as a possible extension of the classical algorithm to the quantum world \cite{fujii2017harnessing,nakajima2019boosting,chen2019learning,ghosh2019quantum}. The first work \cite{fujii2017harnessing} proposed networks of quantum spins as the reservoir, exploiting the large number of degrees of freedom that they can provide with a few elements. Since then, several works have explored the spin-based implementation \cite{nakajima2019boosting,chen2019learning}. 

QRC bridges the fields of neuromorphic computing and quantum information technologies \cite{markovic2020quantum}. The former is inspired by the computation of the brain to reduce energy consumption and time of computation, while the latter harnesses properties of quantum systems, such as entanglement or coherence, to tackle some class of problems that otherwise would be hard to solve in the classical domain \cite{nielsen2010quantum,ladd2010quantum}.
Quantum effects can be exploited in quantum metrology, secure quantum communications, quantum simulations, and quantum computation \cite{acin2018quantum}, being the latter the most relevant technology here. As opposed to quantum circuit models, which implement digital gate-based quantum computing schemes
\cite{nielsen2010quantum},  QRC belongs to the broad line of quantum analog computation, as quantum simulations or quantum annealers, where  the complex real-time dynamics of quantum systems is exploited for  different tasks.

In this work, we consider a quantum network of randomly coupled spins for reservoir computing as in the proposals of Refs. \cite{fujii2017harnessing,nakajima2019boosting,chen2019learning}.
Our objective is to determine the task-independent capabilities of the spin-based QRC system through the Information Processing Capacity (IPC) \cite{dambre2012information}. The IPC is a generalization of the linear memory capacity \cite{jaeger2002short} and quantifies the different degrees of nonlinearity that a dynamical system can reproduce. Equipped with this, we find conditions for an optimum input driving and provide different alternatives for the choice of the system's observables used for the readout. The presented analysis encompasses both time multiplexing as well as different measured spin projections and correlations in order to establish the best performance of a quantum reservoir of interacting spins.

\section{Methods}

\subsection{Reservoir Computing}

Dynamical systems can be exploited to solve temporal tasks. By a temporal task, it is meant to learn a function of a time series (called input sequence) that is fed into the system. If the proposed task is to approximate a nonlinear function of the input, we say that it is then a nonlinear temporal task. Some examples of this are speech recognition \cite{brunner2013parallel,triefenbach2014large,larger2017high} and chaotic time series prediction \cite{jaeger2004harnessing,soriano2014delay,pathak2018model}. The field of Reservoir Computing (RC) harnesses dynamical systems to tackle this kind of problems from a machine learning perspective, applying in general, a supervised learning scheme. The standard algorithm of RC can be divided in the following three steps:
\begin{enumerate}
    \item Feeding an input into the dynamical system, also known as reservoir layer, through some of the variables of it. 
    \item Evolving the natural dynamics of the reservoir layer under the input driving.
    \item Extraction of the information from the reservoir layer, using all or some of the degrees of freedom of the reservoir, via an output layer.
\end{enumerate}

The first step needs to be specified according to the dynamical system in use, so we will explain the input encoding for the spin-based QRC model in the next Section. Nevertheless, in a general setting, we can describe the processing and extraction of information with two maps. First, the reservoir map $T: \mathbb{R}^N\times \mathbb{R}^n\rightarrow \mathbb{R}^N$ with $N,n \in \mathbb{N}$ the dimensions of reservoir's state space and input respectively. Second, the readout map $o: \mathbb{R}^N\rightarrow \mathbb{R}$. In this setting, the reservoir computing paradigm can be represented as follows \cite{grigoryeva2018echo}:
\begin{equation}\label{Eq:RC}
\left\{\begin{matrix}
 \textbf{x}_{k}&=&T(\textbf{x}_{k-1},\textbf{s}_k)
\\
 y_k&=&o(\textbf{x}_{k}),
\end{matrix}\right.
\end{equation}
where $\textbf{x}_{k}\in \mathbb{R}^N$ is the state vector of the dynamical system at time $k\in \mathbb{Z}$, and $\textbf{s}_k\in \mathbb{R}^n$ is the input vector that belongs to an infinite discrete-time input sequence $\textbf{s}=(\dots,\textbf{s}_{-1},\textbf{s}_{0},\textbf{s}_{1},\dots)\in (\mathbb{R}^n)^{\mathbb{Z}}$ (in other words, $\textbf{s}$ is the list of inputs in time). $y_k\in \mathbb{R}$ is the output signal at time $k$. 

Reservoir and output in Eq.~(\ref{Eq:RC}) play different roles. The former is often selected such that it provides a nonlinear transformation with respect to an input in the present time and to past inputs as well. In this way, most of the computational cost can be outsourced to the reservoir while the readout function is kept as simple as possible. Thanks to this simplification, the readout can be taken as a linear combination of the reservoir states:
\begin{equation}\label{Eq:output}
    y_k=\textbf{w}^{\top}\textbf{x}_{k},
\end{equation}
where $\textbf{w}$ represents the output weights that are trained in the supervised learning scheme. It has been extensively shown that it is enough to train only these weights, keeping the parameters of the dynamical system fixed from the beginning of the learning process \cite{lukovsevivcius2009reservoir,grigoryeva2018echo}.
\begin{figure}[htb]
\begin{center}
\includegraphics[width=\columnwidth]{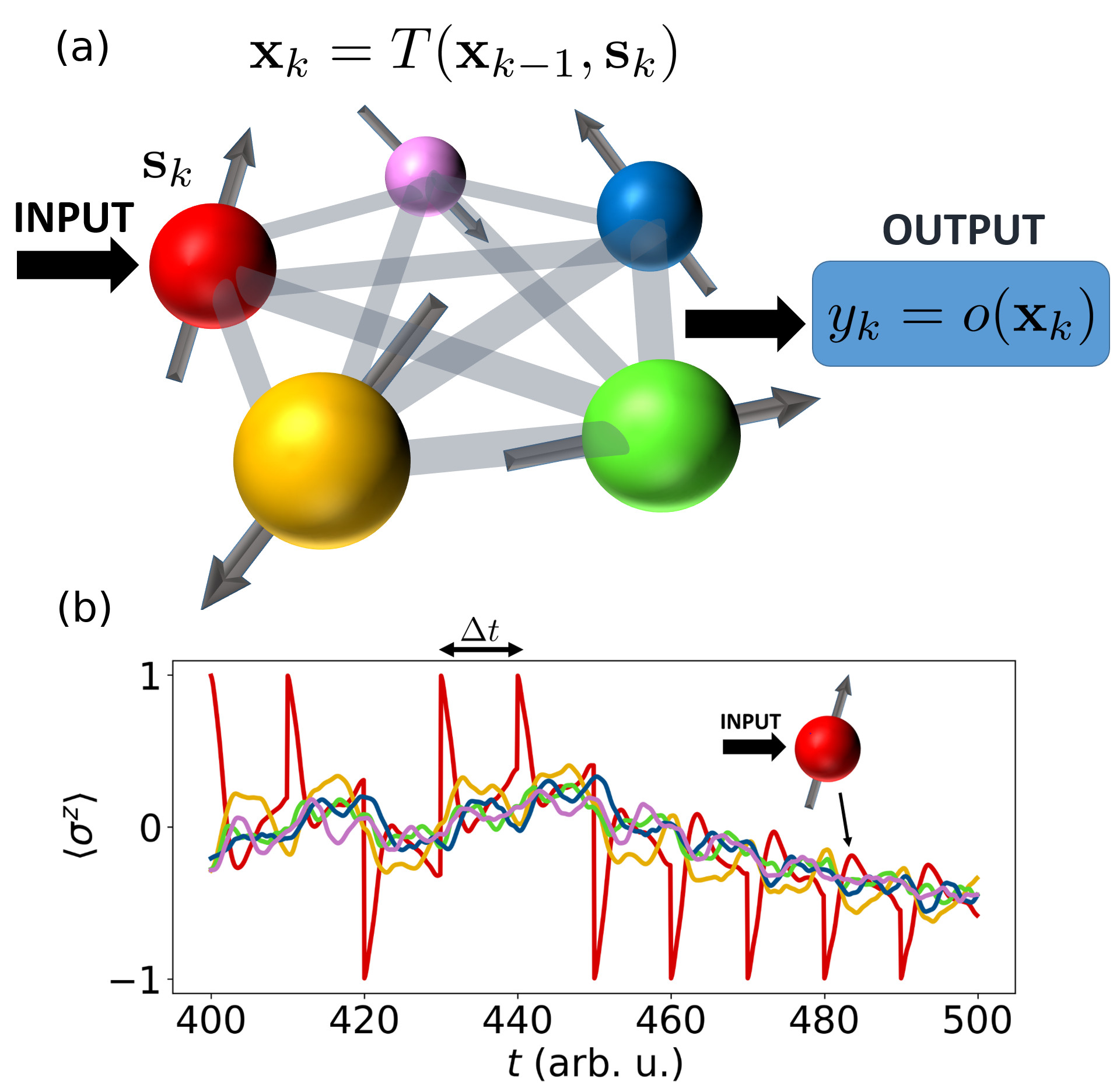}
\caption{(a) Schematic representation of a reservoir computing system with a network of 5 quantum spins. (b) Example of the dynamics of the observables $\braket{\sigma^z_i}$ for a random network of 5 spins with parameters $\Delta t=10$, $h=1$ and $J_s=1$. It can be seen that the spins 2, 3, 4 and 5 are driven by the dynamics of spin 1, plotted with the red line.}\label{Fig:1}
\end{center}
\end{figure}
The training is the part of the algorithm where the weights $\textbf{w}$ are adjusted to solve a task by minimizing the error. We will consider here the mean square error
\begin{equation}\label{Eq:MSE}
MSE_L(\textbf{y},\bar{\textbf{y}})=\frac{1}{L}\sum^L_{k=1}(y_k-\bar{y}_k)^2,
\end{equation}
where $\bar{y}_k$ is the function of the input that we want to reproduce, also known as target. A usual approach is to use a linear regression for vector $\textbf{y}=X\textbf{w}$  with respect to $\bar{\textbf{y}}$, where $X=\left(\textbf{x}_1,\dots,\textbf{x}_k\right)^\top$ is a $L\times(N+1)$ matrix that collects the reservoir states at different times. $L$ is the length of the input sequence and $N$ is the  number of variables, i.e. observables, we use from the dynamical system (we add a constant bias term $x_{k,N+1}=1$ for an optimal training).
\subsection{Model}

As discussed in the previous section, dynamical systems can be exploited in the RC framework. In particular, Quantum Reservoir Computing (QRC) was introduced as the possibility of employing quantum dynamical systems \cite{fujii2017harnessing}. In this approach, we use a quantum system for obtaining the reservoir as defined in Eq.~(\ref{Eq:RC}), while the system evolves under the usual expressions of quantum mechanics. For a closed system, the dynamical evolution of the spin-based QRC model can be written as
\begin{equation}\label{Eq:U}
\rho(k\Delta t)=e^{-iH\Delta t}\rho[(k-1)\Delta t]e^{iH\Delta t},
\end{equation}
 where $\rho(k\Delta t)$ is the density matrix of the quantum system after $k$ time steps and $e^{-iH\Delta t}$ is the unitary evolution given by Hamiltonian $H$ for a time interval $\Delta t$. Here, we will deal with networks of qubits represented in Fig.~\ref{Fig:1} (a), as in Ref. \cite{fujii2017harnessing}. The Hamiltonian we use is known as the transverse-field Ising model and can be written as
\begin{equation}\label{Eq:H}
    H=\sum^N_{i>j=1}J_{ij}\sigma_i^x\sigma_j^x+h\sum^N_{i=1}\sigma_i^z,
\end{equation}
where $N$ is the number of qubits, $h$ is their natural frequency, $\sigma^{a}_i$ with $a=x,y,z$ are the usual Pauli matrices and $J_{ij}$ are the couplings between the qubits. These couplings will be selected at random from a uniform distribution in the interval $[-J_s/2,J_s/2]$. Nonuniform couplings are in general beneficial to explore the whole Hilbert space and then to exploit all degrees of freedom.

Once we have defined the dynamical system, we can introduce the encoding of the input signal  into the system, i.e. $s_k$, every time step. From now on, we will work only with a scalar input. We will make use of the Completely Positive Trace-Preserving (CPTP) map in which the classical information $s_k$ is fed always to the same qubit \cite{fujii2017harnessing}. Let us name it qubit one. The state of this qubit is reinitialized every $\Delta t$ with the state
\begin{equation}\label{Eq:S}
\ket{\psi_{s_k}}=\sqrt{1-s_k}\ket{0}+\sqrt{s_k}\ket{1}, ~~~s_k\in [0,1],
\end{equation}
which can be represented with the density matrix $\rho_1=\ket{\psi_{s_k}}\bra{\psi_{s_k}}$. The new density matrix of the whole system is given by the CPTP map
\begin{equation}\label{Eq:U1}
\rho'(k\Delta t)=\rho_1 \otimes \text{Tr}_1\left\{\rho[(k-1)\Delta t]\right\},
\end{equation}
where $\text{Tr}_1\left\{\cdot\right\}$ denotes the partial trace performed over the first qubit.
Gathering the maps in Eqs.~(\ref{Eq:U}) and (\ref{Eq:U1}), the first two steps of the algorithm, i.e. feeding an input and processing it, can be summarized in the total map 
\begin{equation}\label{eq:map}
\rho(k\Delta t)=e^{-iH\Delta t}\rho_1 \otimes \text{Tr}_1\left\{\rho[(k-1)\Delta t]\right\}e^{iH\Delta t}.
\end{equation}
In order to fully define the spin-based QRC, we still lack reading out the information of the system. 
The way in which we can extract information from a quantum system is by measuring it. The corresponding observables are the elements of the N-qubit Bloch vector and can be calculated as
\begin{equation}\label{Eq:nonl}
x_j(k\Delta t)=\text{Tr}[B_j \rho(k\Delta t)]=\braket{B_j},
\end{equation}
where $B_j$ is a product of $N$ Pauli operators taken from $\{I,\sigma^x,\sigma^y,\sigma^z\}$. For example, 
the projection observable of the first spin over the $z$-axis is given by $B=\sigma^z_1\otimes I_2\otimes \cdots \otimes I_N$.
In Fig.~\ref{Fig:1} (b), we show the dynamics of observables $\braket{\sigma^z_i}$ when they are driven by the input.

The observables are the variables of our RC system, as they encode the full quantum state. For $N$ spins, the number of linearly independent variables that we can obtain is as large as $4^N-1$ (the $-1$ comes from the normalization). Measuring such a large number of observed variables is in general not practical and a set of observables is selected for the output layer. In a real experiment, the choice of this set could be conditioned from the setup. For most of the numerical experiments, we will take the projections over the $z$-axis of the spins $\braket{\sigma^z}$, in order to compare with the observations reported in \cite{fujii2017harnessing}. However, as we will see in Subsection \ref{Sec:observables}, a key issue is about advantages or limitations when measuring different observables, also signaling qubits correlations.
 
As in \cite{fujii2017harnessing}, the present approach is based on an ensemble quantum system, which contains a large number of copies of the reservoir. An example of this kind of ensemble can be found in NMR spin ensemble system, like the experiment carried out in \cite{negoro2018machine}. This assumption allows us to disregard the back-action after any measurement.

\subsection{Information processing capacity}

It was recently shown that one can estimate the computational capabilities of a dynamical system measuring its Information Processing Capacity (IPC) \cite{dambre2012information}. 
This approach is particularly insightful, as it provides the full task-independent characterization of reservoir computers.
In particular, the authors of Ref. \cite{dambre2012information} demonstrate that the total computational capacity of a dynamical system is bounded by the number of linearly independent variables that we use for the output. In addition, the computed IPC can only saturate the bound when the dynamical system has fading memory, i.e. when the dynamical system dissipates the information about the input after some time.

In general, a given capacity quantifies how well a system reproduces a target function, and it is defined as
\begin{equation}\label{Eq:C}
    C_L(X,\textbf{y})=1-\frac{\text{min}_{\textbf{w}}MSE_L(\textbf{y},\bar{\textbf{y}})}{\braket{\bar{\textbf{y}}^2}_L},
\end{equation}
where we recall that $X$ is the matrix of the reservoir variables at different times, $\textbf{y}$ is the vector of outputs from the reservoir for each input, $\bar{\textbf{y}}$ is the target and $\textbf{w}$ is the vector of weights of the output layer. The mean square error introduced in  Eq.~(\ref{Eq:MSE}) can be obtained from the outputs of the system, for example with a linear regression. The bracket $\braket{\bar{\textbf{y}}^2}_L$ denotes the temporal average over the  target sequence of size $L$.

To compute the total capacity of the spin-based QRC system, we should evaluate all the possible linear and nonlinear functions of the input sequence that our system can approximate.
Following the original theory \cite{dambre2012information}, the chosen target functions need to be orthogonal and will be defined as the product of Legendre polynomials of a given degree $d_i$, such that the sum of the degrees of all the multiplied polynomials add up to a given degree $d$ of nonlinearity. Then, the total capacity can be divided in linear and nonlinear contributions. Besides, the polynomials will be functions of the input at different times in the past. In this way, our targets not only account for nonlinear functions but also temporal maps. 

The target function for a given degree $d$ is:
\begin{equation}\label{Eq:Pol}
    \bar{y}_k=\prod_i \mathcal{P}_{d_i}[\tilde{s}_{k-i}], \quad \sum_{i}d_i=d,
\end{equation}
where the scaled input is uniformly distributed over the interval $[-1,1]$, consistently with the polynomials definition.
Comparing with Eq.~(\ref{Eq:S}), $s_k=(1+\tilde{s}_k)/2$.

In the theoretical framework, the total capacity is saturated when input sequences of infinite length are considered and the contributions of all the possible nonlinearities, i.e. up to an infinite degree, are computed. However, a sufficiently long input sequence and sufficiently high maximum degree give a stable result. In practice, this means input streams of a length $L=10^5$, as in our simulations.
As for degree, we have checked that $d_{\text{max}}=9$ is enough  for our system. Besides, an initial transient must be taken into account to wash out the initial condition of the system, that for us will be $10^4$ inputs. Since individual capacities may be slightly overestimated in the numerical analysis, a threshold is set to truncate the smaller contributions (for a more formal definition of this threshold, see \cite{dambre2012information}).

\section{Numerical results}

\subsection{Influence of the input driving on the fading memory}
\begin{figure}[htb]
\begin{center}
\includegraphics[width=\columnwidth]{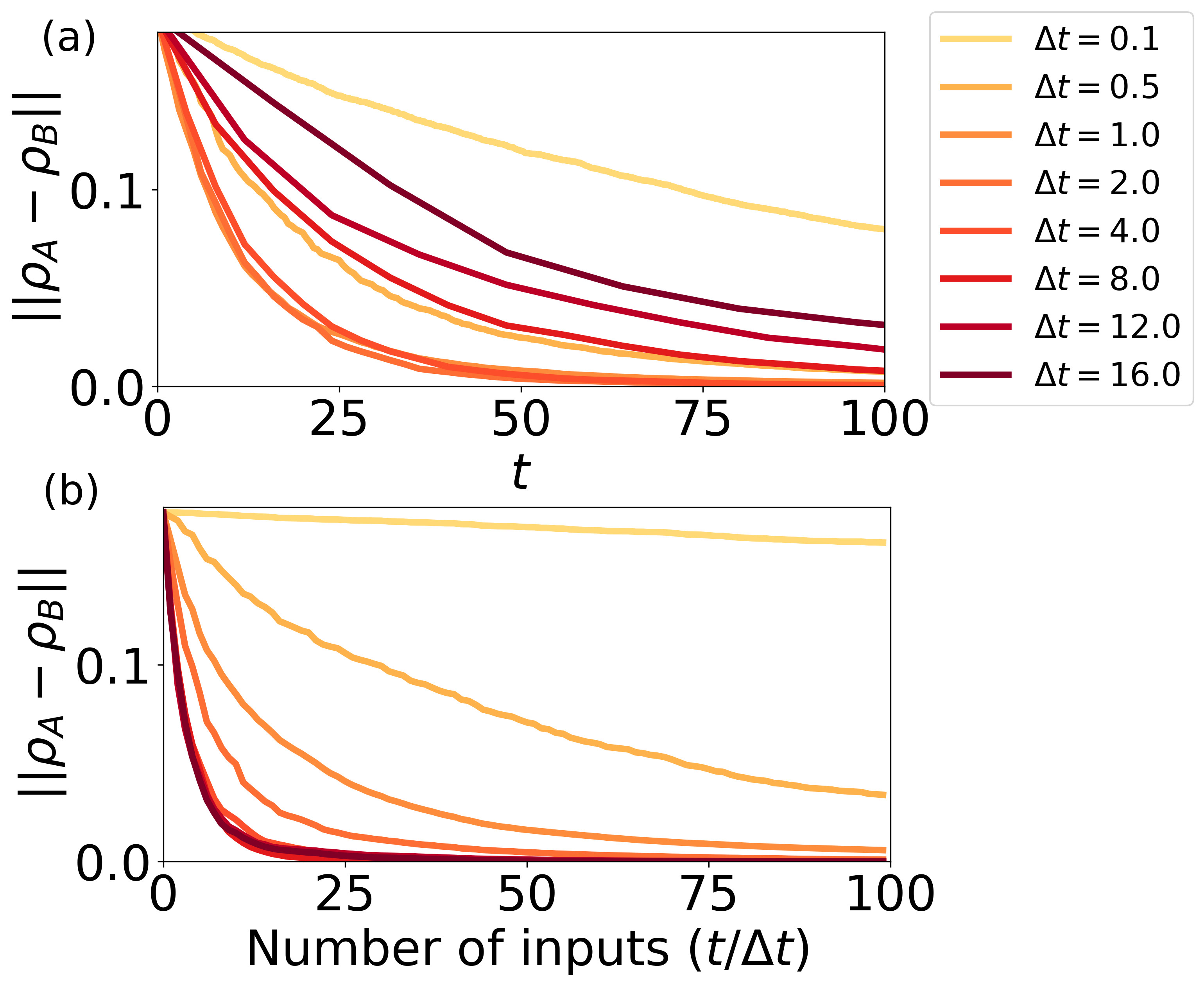}
\caption{Illustration of the convergence of two distant initial states with respect to the same input sequence varying the times $\Delta t$ at which inputs are fed into the system. The system parameters are $N=5$, $h=1$ and $J_s=1$. In (a), the $x$-axis represents physical time in arbitrary units. The number of inputs at a certain time is given by $\text{int}(t/\Delta t)$. In (b), the convergence is shown with respect to the number of inputs.}\label{Fig:2}
\end{center}
\end{figure}

To verify that our system is well behaved, we can start checking if it has fading memory, i.e., if it forgets the initial condition and also how fast it does so. This can be evaluated by simulating the temporal evolution of the system for the same input sequence but starting from two different distant initial states of the quantum spins network, $\rho_A$ and $\rho_B$. We will focus on the influence of the time between inputs $\Delta t$ since this parameter is directly involved in the input driving, controlling the interaction time of the system's Hamiltonian for every input. The other parameters are fixed as $h=1$ and $J_s=1$,
and the number of spins is $N=5$, here and in the following, unless otherwise stated. The influence of parameters $h$ and $J_s$ on the IPC is explored in Appendix A, which also includes a justification for this choice. The corresponding numerical results are shown in Fig.~\ref{Fig:2}. 

Fig.~\ref{Fig:2} (a) represents the distance between the density matrices obtained from two different initial conditions at different instants of time $t$.
This distance follows from the Frobenius norm defined as $||A||=\sqrt{\text{Tr}(A^\dagger A)}$ where $A$ is an arbitrary complex matrix and $A^\dagger$ is its conjugate transpose. We find that the time to reach convergence, i.e., a vanishing difference of states, depends on the value of $\Delta t$. The fastest convergence is found for an intermediate value of $\Delta t\sim1$, while slow convergence occurs for both small and large values of $\Delta t$.
Actually, there are two relevant factors influencing convergence. First, the frequency at which the system is driven away from its free evolution due to input injection that comes with an associated information erasure (partial trace in Eq.~\ref{eq:map}). And second, the free network relaxation to a state that, with the erasure of information of the partial trace, will tend after a certain transient to be independent of the initial conditions. Too frequent injections almost freeze the system while increasing $\Delta t$ the system has enough time to relax. We observe indeed an interplay between the time we wait for feeding the input, $\Delta t$, and the number of inputs we introduce during the evolution $t$.
The effect of the number of inputs, defined as $t/\Delta t$, is addressed in  Fig.~\ref{Fig:2} (b). 
In this representation, we find that all curves for the convergence collapse for $\Delta t\gtrsim4$.
In contrast, the number of inputs needed to reach convergence increases for $\Delta t\lesssim4$, as expected since the system does not have time to significantly evolve.
The results in Fig.~\ref{Fig:2} illustrate that values of the time between inputs in the range $\Delta t\in[1,4]$ provide a compromise between a fast convergence and a moderate number of inputs to reach convergence, already attained at time $t\sim 50$. 
We note that, while we have never observed convergence to be lost with the chosen parameter values, convergence may depend on the particular instance of an input sequence.

\subsection{Influence of the input driving and network size on the information processing capacity}
\begin{figure}[h!]
\begin{center}
\includegraphics[width=\columnwidth]{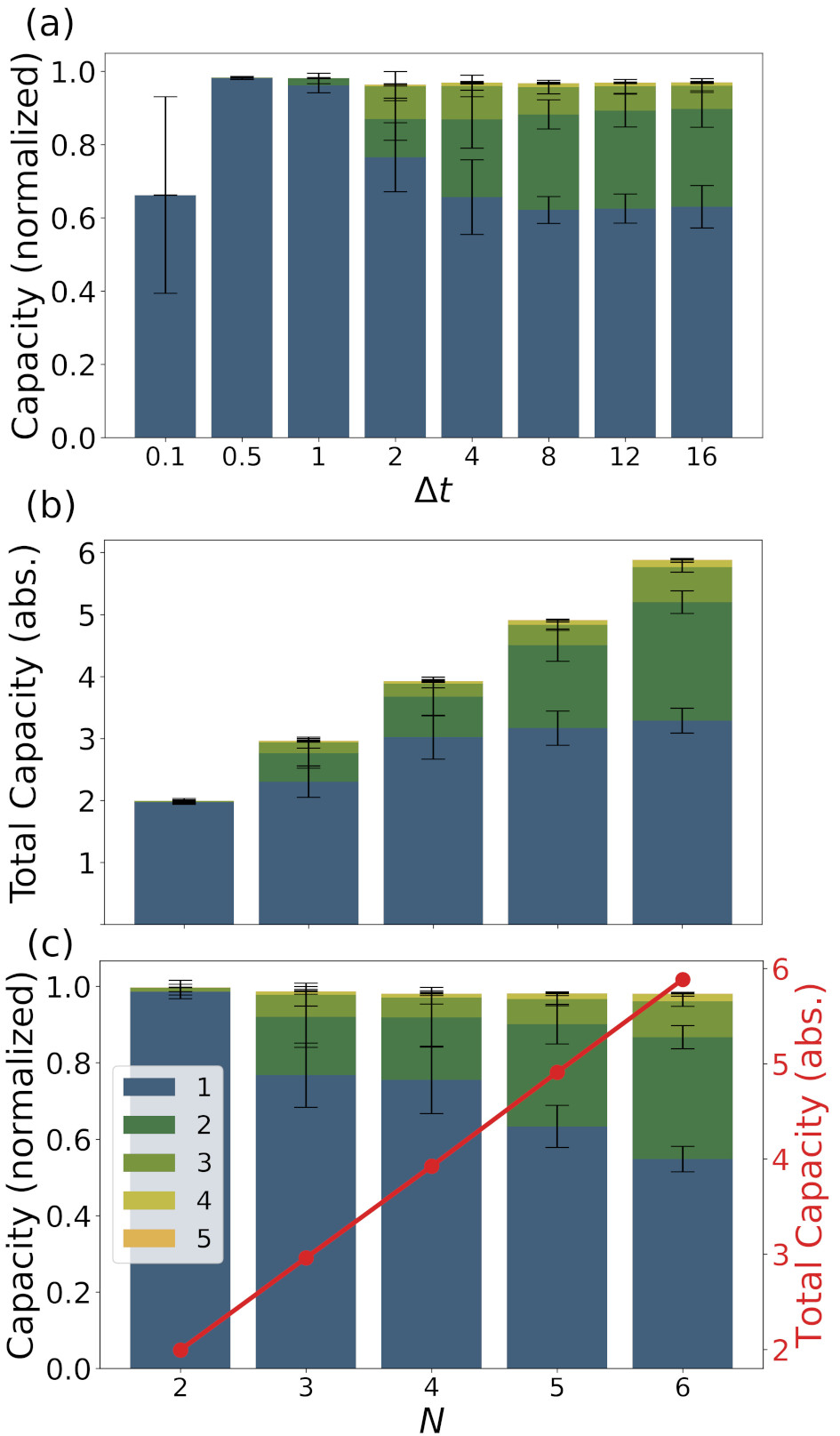}
\caption{(a) IPC versus time between inputs $\Delta t$; in (b) and (c) respect to the number of spins $N$. The parameters in (a) are $h=1$, $J_s=1$ and $N=5$, while in (b) and (c) we have used the same $h$ and $J_s$ with $\Delta t=10$. In (b), we represent the IPC without normalization to appreciate the growth of nonlinearities. In (c) we normalize the IPC respect to the theoretical maximum capacity (number of output variables, here $N$). In this last plot we include the behaviour of the total capacity in absolute value (red line) to stress that the capacity of the system increases with the number of observables in the output. The error bars of the plots correspond to the standard deviation over 10 realizations. }\label{Fig:3}
\end{center}
\end{figure}
Now we focus on the influence of the input driving, through $\Delta t$, on the IPC evaluated for the first time for spin-based QRC.
The corresponding numerical results are shown in Fig.~\ref{Fig:3}, together with the influence of the number of reservoir spins $N$ on the IPC.
In the following, the values of the IPC are usually given in normalized units, with the normalization factor being the respective number of output variables (observables), i.e., the number of trained output weights.
For the sake of clarity, also the absolute capacity is shown when relevant. Insight is gained from the displayed contributions to the IPC corresponding to different degrees, plotted in different colors.
Fig.~\ref{Fig:3} (a) shows the IPC as a function of the time between inputs $\Delta t$, with the number of observables ($\braket{\sigma^z_i}$) being $N=5$.

For small and intermediate values of $\Delta t$, i.e. $\Delta t\lesssim 1$, we observe in Fig.~\ref{Fig:3} (a) that the maximum capacity is reached and that the contribution to the IPC is mostly linear, i.e. $d=1$.
Nonlinear contributions to the IPC appear as $\Delta t$ increases, mostly dominated by degrees $d=2$ and $d=3$.
Finally, we find that the ratio of linear and nonlinear contributions to the IPC remain constant when $\Delta t\gtrsim4$.

To a large extent, we can relate the trends of the IPC in Fig.~\ref{Fig:3} (a) to the results presented in Figs.~\ref{Fig:2} (a) and (b). For small $\Delta t$, the convergence becomes extremely slow and the distance between the states originating from two different initial conditions is barely affected by new inputs. Thus, the maximum IPC is not reached.
For $\Delta t\sim1$, the system shows a fast convergence but the IPC indicates that the interactions between the inputs and the reservoir's observables remain mostly linear.
For $\Delta t\gtrsim4$, the number of inputs to reach convergence become the dominant parameter and it remains constant. 
As a result, the capacity of the system no longer changes.
The results in Fig.~\ref{Fig:3} (a) illustrate that the spin-based QRC system possesses nonlinear capacity contributions as long as the interaction time between inputs is sufficiently long.
These results help us to decide the value of $\Delta t$ for the following numerical simulations.
We choose $\Delta t=10$ as a representative value of the rich system's dynamical response, showing both linear and nonlinear contributions to the IPC. This value is well established within the saturation region of the maximum capacity for this set of parameters and observables.

Finally, Figs.~\ref{Fig:3} (b) and (c) show the IPC as a function of the number of spins in the system, $N$. For the selected parameters, the total memory is always saturated and increases linearly with the system size, corresponding to the number of spins/$z$-observables. Interestingly, nonlinear contributions show up when we increase the number of spins. 
This indicates that the system complexity increases with increasing network sizes.
From now on, whenever we want to show the total capacity, it will be contained in the right axis of the figure of interest, like in Fig.~\ref{Fig:3} (c).
With respect to $N$, we take $N=5$ as our benchmark for the following numerical simulations in order to keep a reasonable computational time.
This assessment allows to identify a good compromise for the injection speed and size choice for the performance of the QRC with Ising network of spins.

\subsection{Influence of the observables on the information processing capacity: virtual nodes }\label{Sec:observablesV}

\begin{figure}[htb]
\begin{center}
\includegraphics[width=\columnwidth]{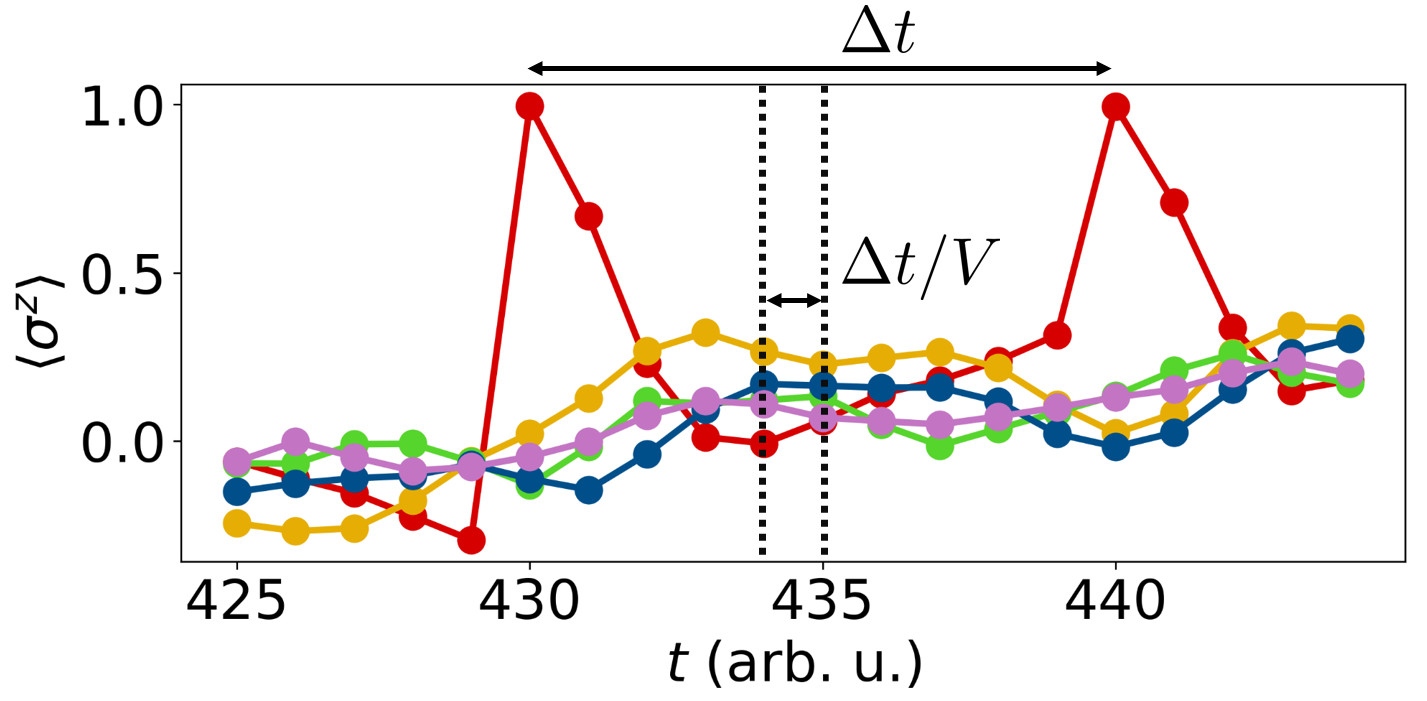}
\caption{Representation of $V$ additional measurement intervals on the dynamics of $\braket{\sigma^z_i}$ for a system with $N=5$ spins. Between two different inputs, the dynamics is sampled in $V$ steps with time $\Delta t/V$ between them. The red line corresponds to the spin labeled as 1 (with the input) and the rest of colors correspond to spins from 2 to 5.}\label{Fig:4}
\end{center}
\end{figure}

So far, we have limited our  observables to a small set, the projections over the $z$-axis $\braket{\sigma^z_i}$ for $i=1,\dots, N$. However, this is a small portion of information compared to the large number of degrees of freedom that our system can provide and the rich dynamics that they present. A proposal introduced in classical \cite{appeltant2011information} and quantum \cite{fujii2017harnessing} versions of reservoir computing tries to exploit this last point. It is known as time multiplexing and consists in sampling the observables at smaller time intervals between two different inputs. In this manner, we take more than one snapshot of the dynamics, obtaining more information that was hidden to us with only one measurement between inputs. The time multiplexing scheme used here is represented in Fig.~\ref{Fig:4}. Having already $N$ spins, we will increase the number of output variables up to $NV$, where $V$ is the number of subdivided time intervals. These additional variables are often referred to as virtual nodes \cite{appeltant2011information,fujii2017harnessing}.

We first need to derive a formal definition of the additional observables when time multiplexing is considered.
Once we have introduced the input $s_k$ into the system, we evolve the density matrix for a time interval $v\Delta t/V$ with $1\leq v\leq V$ using the following map:
\begin{equation}
\begin{split}
&\rho[(k-1)\Delta t+v\Delta t/V]=\\
&e^{-iHv\Delta t/V}\rho_1 \otimes\text{Tr}_1\left\{\rho[(k-1)\Delta t]\right\}e^{iHv\Delta t/V},
\end{split}
\end{equation}
getting the state of the system at different times between inputs.
Then, the value of the observables can be computed also at different times
\begin{equation}
\begin{split}
&x_j[(k-1)\Delta t+v\Delta t/V]=\\
&\text{Tr}\left\{B_j \rho[(k-1)\Delta t+v\Delta t/V]\right\}.
\end{split}
\end{equation}

\begin{figure}[htb]
\begin{center}
\includegraphics[width=\columnwidth]{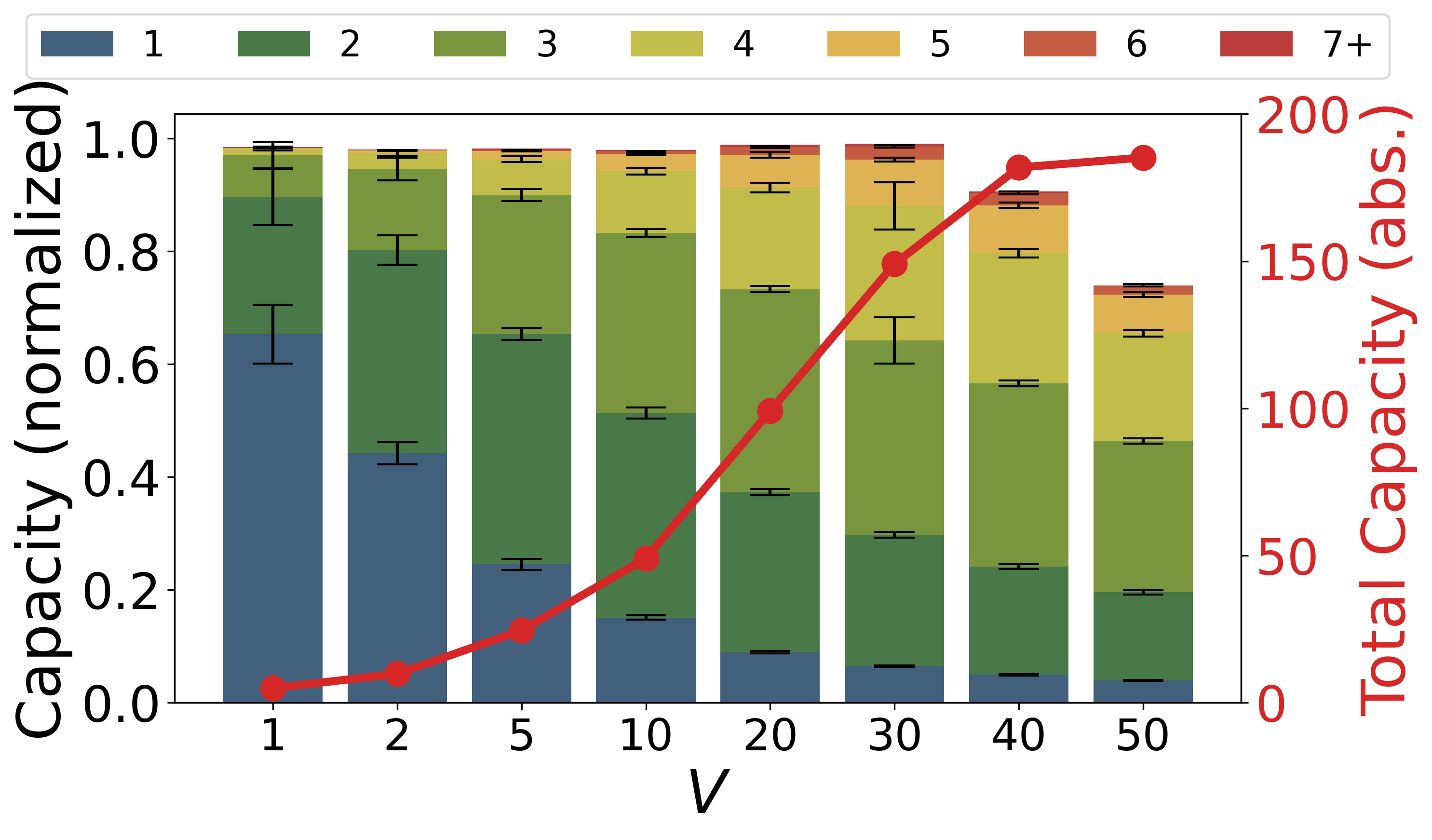}
\caption{IPC versus number of time-multiplexed segments $V$. The parameters are $N=5$, $\Delta t=10$, $h=1$ and $J_s=1$. The red line represents the total capacity in absolute value. Notice that the normalization factor in this plot is $N$ times $V$. The error bars of the plots correspond to the standard deviation over 10 realizations.}\label{Fig:5}
\end{center}
\end{figure}

Fig.~\ref{Fig:5} illustrates the influence of considering virtual nodes over the processing capacity of the system. Increasing the number of output variables, we take advantage of the rich dynamics of the observables and increase the total
capacity of the system (in absolute value, see red line in Fig.~\ref{Fig:5}), up to a saturation when virtual nodes become redundant.
However, for high enough values of $V$, we observe that the normalized maximum capacity is not reached. For high $V$, the time interval $\Delta t/V$ between virtual nodes becomes too small to provide a significant temporal evolution of the observables. 
In fact,  for small values of $\Delta t/V$, the exponential in the Hamiltonian dynamics between virtual nodes can be approximated at the first order, making the network evolution nearly linear.
Another interesting finding in Fig.~\ref{Fig:5} is that, by considering additional virtual nodes, higher-order degree nonlinearities become available.
These results suggest that combining the observables for different spins at different times unlocks the dynamical complexity that is otherwise inaccessible in the QRC-based spin system.
Further research can be focused on the origin of these nonlinearities.

\begin{figure}[htb]
\begin{center}
\includegraphics[width=\columnwidth]{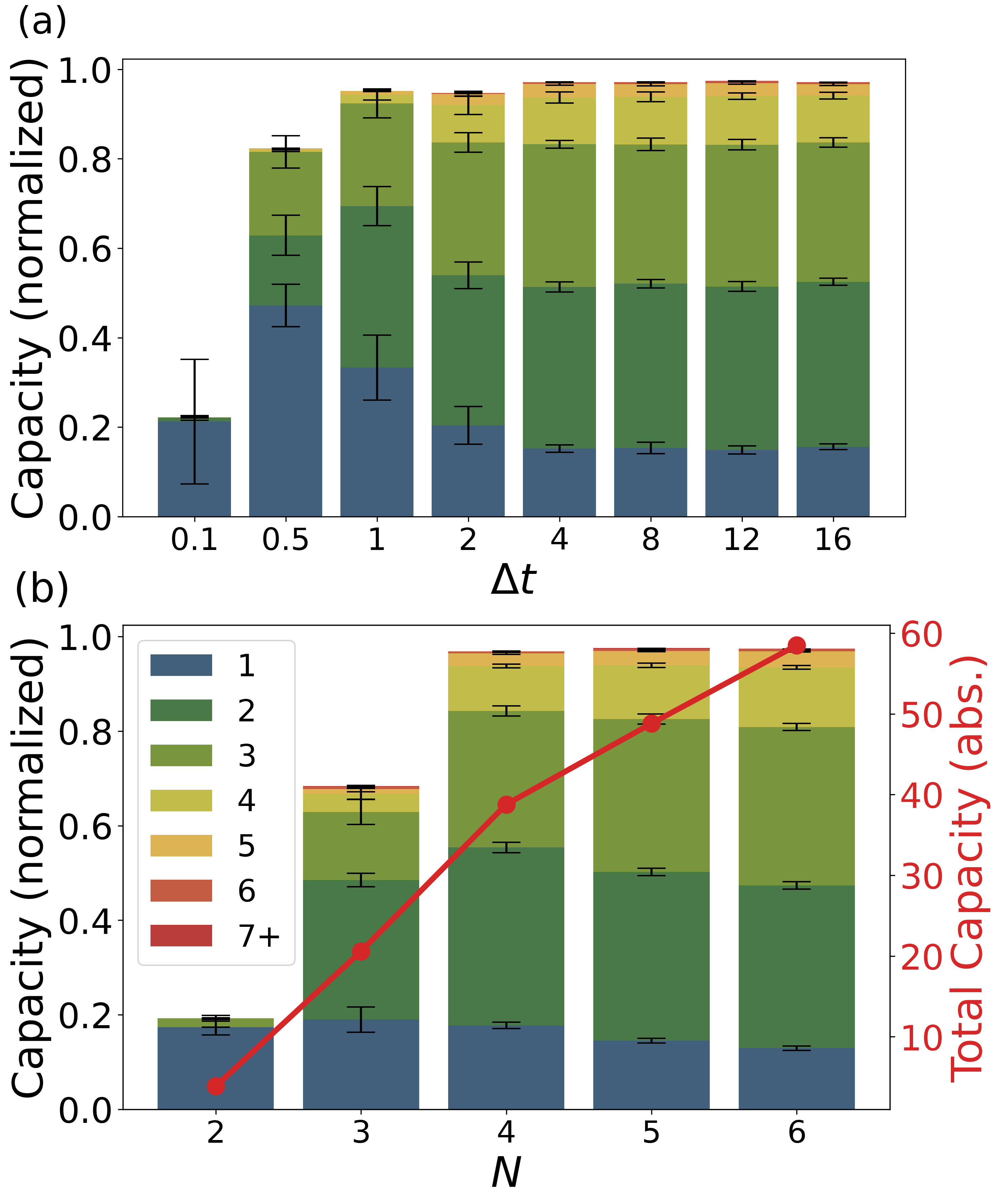}
\caption{IPC versus (a) time between inputs $\Delta t$ and (b) number of spins $N$, both with a number of time-multiplexed segments $V=10$. The parameters in (a) are $h=1$, $J_s=1$ and $N=5$, while in (b) we have used the same $h$ and $J_s$ with $\Delta t=10$. The red line in (b) denotes the total capacity in absolute value. Notice that the normalization factor in this plot is $N$ times $V$. The error bars of the plots correspond to the standard deviation over 10 realizations.}\label{Fig:6}
\end{center}
\end{figure}

We have shown in Fig.~\ref{Fig:5} that introducing virtual nodes can have a large impact over the IPC, even if the remaining system parameters remain unchanged. Now, we evaluate the impact of having virtual nodes on the IPC as a function of other system parameters, in an analogous manner to the numerical results leading to Fig.~\ref{Fig:3}.
By setting an intermediate number of time-multiplexed segments, $V=10$, we proceed to illustrate the behaviour of the IPC with respect to $\Delta t$ and $N$ in Fig.~\ref{Fig:6}, in which the number of output variables is $10N$.
The results in Fig.~\ref{Fig:6} (a) show that the linear and nonlinear contributions to the IPC again remain constant when $\Delta t\gtrsim4$.
We find that the maximum normalized capacity is in this case reached when $\Delta t>2$.
In a relative sense, the nonlinear contributions become more apparent when virtual nodes are considered, as compared with the results in Fig.~\ref{Fig:3}.
In addition, we observe a more pronounced decrease of the normalized capacity in Fig.~\ref{Fig:6} for small values of $\Delta t$ with respect to Fig.~\ref{Fig:3}.

The fading memory of the system is certainly not affected by the consideration of virtual nodes since the convergence of the spin-based QRC system is independent of $V$. 
We find that the loss of capacity for small $\Delta t$ in Fig.~\ref{Fig:6} (a) comes from the similarity between output variables, with the virtual nodes becoming linearly dependent. 
Finally, we show in Fig.~\ref{Fig:6} (b) that the maximum normalized capacity  is only reached for $N\geqslant4$.
For small system sizes, the increase in the number of output variables appears to fail to provide information due to the fact that some of these output variables are not linearly independent. 
Summarizing, there is a limit to the total achievable capacity for an increasing number of virtual nodes given a fixed system size and choice of observables, being the major advantage of introducing virtual nodes the possibility to access nonlinear memory.
\begin{figure*}[htb]
\begin{center}
\includegraphics[width=\textwidth]{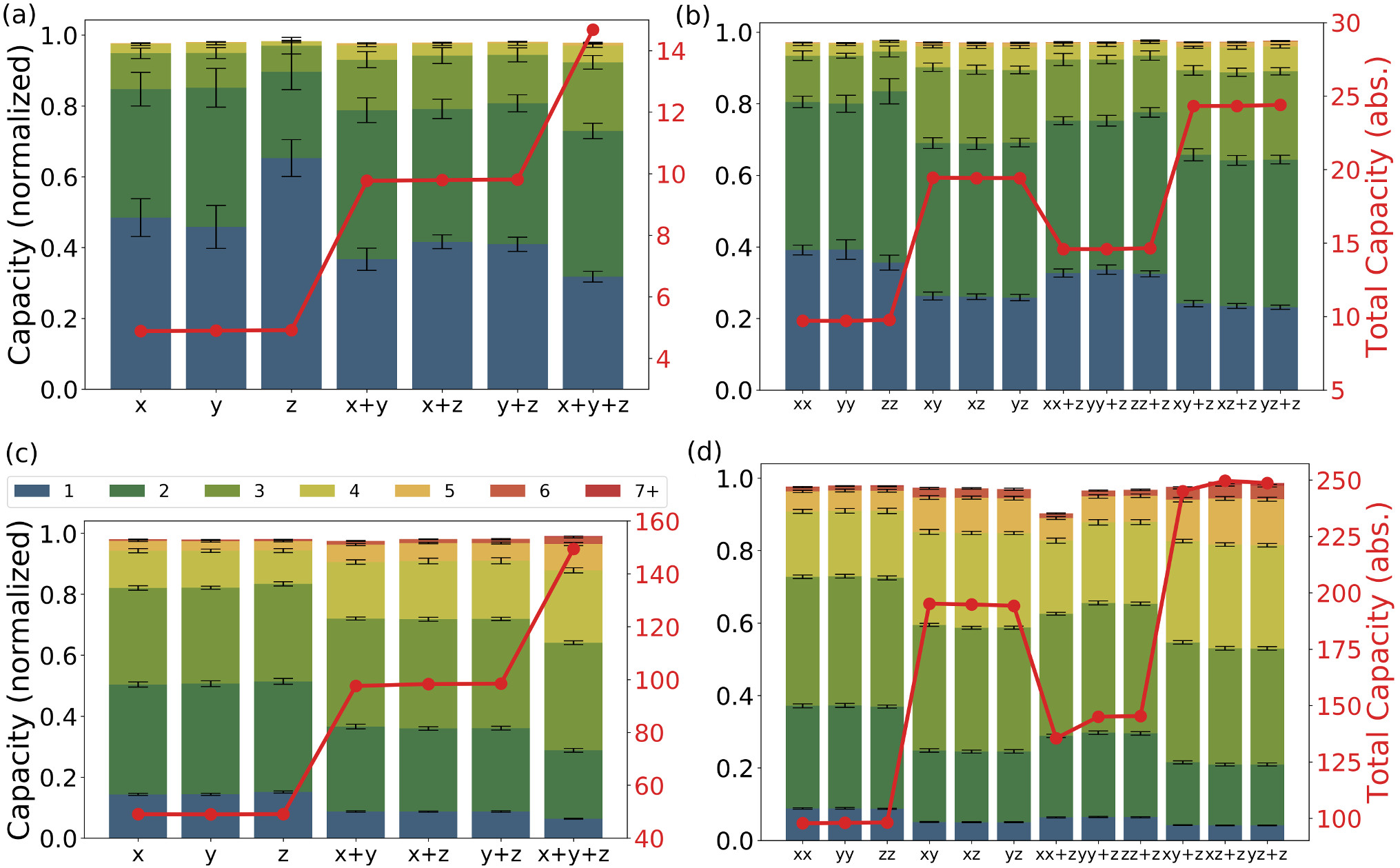}
\caption{IPC versus different sets of observables. The parameters are $N=5$, $\Delta t=10$, $h=1$ and $J_s=1$. The red lines represent the total capacity in absolute value. The top row ((a) and (b)) corresponds to $V=1$ while the bottom row ((c) and (d)) corresponds to $V=10$. 
Notice that the normalization factor in this plot is $V$ times the number of observables (see the main text for details). The error bars of the plots correspond to the standard deviation over 10 realizations.}\label{Fig:7}
\end{center}
\end{figure*}

\subsection{Influence of the observables on the information processing capacity: different projection directions and correlations}\label{Sec:observables}

 We initially restricted ourselves to use the projections $\braket{\sigma^z_i}$ as our output variables. One may wonder if this is a preferred direction for the QRC performance. Furthermore,
since these are just a few number of observables, $N$, with respect to the total number of degrees of freedom of the spin-based QRC system, it is interesting to explore the performance for other possible choices of the observables, including correlations.
Fig.~\ref{Fig:7} represents the IPC for different sets of observables.
In the top row, Fig.~\ref{Fig:7} (a) shows the IPC of the three different projections of the spin ($x$, $y$ and $z$) and the contributions of their combinations. By combinations, we mean to use, for instance, together the observables of $\braket{\sigma^x_i}$ and $\braket{\sigma^y_i}$ in the output layer, having a total amount of 10 observables ($2NV$, with $N=5$ and $V=1$) and labeled in the plot as x+y. Fig.~\ref{Fig:7} (b) shows the same characterization but for correlations of the form $\braket{\sigma^a_i\sigma^b_j}$, with $a,b=x,y,z$ and $i\neq j$.
At a first sight, the different combinations of observables shown in Figs.~\ref{Fig:7} (a) and (b) do not show a large qualitative difference in normalized capacities. Still, we note
a slightly larger linear memory when considering spin projections in the $z$ direction. Higher total capacities are achieved considering also correlation measurements, as shown in Fig.~\ref{Fig:7} (b). For example, considering xy+z, xz+z and yz+z leads to a total capacity of 25 
(20 for the correlations of $\braket{\sigma^a_i\sigma^b_j}$ and 5 for $\braket{\sigma^a_i}$), while it is only 15 for x+y+z in Fig.~\ref{Fig:7} (a). The available number of output variables for each combination of observables is indicated by the red lines.

The bottom row of Fig.~\ref{Fig:7} provides insights on the role of the observables when considering both virtual nodes and different measured quantities. Figs.~\ref{Fig:7} (c) and (d) represent the IPC for the same sets of observables as in Figs.~\ref{Fig:7} (a) and (b) but using $V=10$ virtual nodes. As shown in Fig.~\ref{Fig:7} (c), the total capacity saturates the bound for the simplest sets of observables and their combinations. The combinations of observables bring additional nonlinear memory contributions while the linear memory remains roughly constant.
Furthermore, using correlations as observables, as is done in Fig.~\ref{Fig:7} (d), we can sustain the total capacity up to the saturation level as well, even with the consideration of virtual nodes and different combinations of observables. In this way, we can obtain a larger total capacity of the system with respect to the cases in Fig.~\ref{Fig:7} (c). For instance, a bar like xy in Fig.~\ref{Fig:7} (d) has a total capacity of 200 
(20 times $V$), while none of the bars in Fig.~\ref{Fig:7} (c) has a total capacity larger than 150.
The large total capacity reached in Fig. 7 (d) illustrates that the spin-based QRC system has a large number of available degrees of freedom even if the system is made of only a few spins, here $N=5$.

\section{Conclusion and Future Work}

Quantum reservoir computing is a new promising line in the already well established field of reservoir computing. Recent theoretical proposals of quantum reservoir networks of spins are leading to preliminary proof-of-principle experimental realizations \cite{negoro2018machine,chen2020temporal}.
In addition, extension of QRC to continuous variable systems \cite{nokkala2020gaussian} or to layered reservoir structures \cite{keuninckx2017real,gallicchio2017echo} offers promising perspectives.

In this work, we have evaluated the performance of the spin-based QRC model proposed in \cite{fujii2017harnessing} with respect to the Information Processing Capacity measure \cite{dambre2012information}. This task-independent measure gathers all the possible contributions that a dynamical system can provide to reproduce linear and nonlinear functions of an input sequence. The IPC allows to quantify these contributions and test whether the system under study is working in proper operational conditions, where the total capacity is maximized. 

The complex dynamics of the spin network has been characterized to establish the fading memory character of the reservoir. The convergence of the system when starting from different initial conditions is strongly influenced by the frequency of input injection. We have identified competing mechanisms of information erasure and system driving, leading to an optimum value of the input injection periodicity ($\Delta t$) so that the system can converge faster. 
In particular, this optimum value corresponds to the thermalization time of local observables, as will be discussed elsewhere \cite{Rodrigo}.
Regarding the IPC, $\Delta t$ sets a threshold for which the total capacity is saturated, finding a stable distribution of linear and nonlinear memories for the higher values. These findings set the conditions to estimate the potential operating speed of spin-based QRC systems in experimental realizations.

The second main result of this work is the identification of the importance of the reservoir and output sizes on the IPC through either virtual nodes and observables. We have explicitly exploited the large dimensionality of spin-based QRC systems and provided ways to access the corresponding available capacity. On the one hand, we saw that exploiting the rich dynamics of observables through time multiplexing can increase significantly the  memory capacity of the system, without compromising the saturation of the total capacity. This multiplexing reaches its limit when $\Delta t/V$ is so small that virtual nodes are not linearly independent anymore. On the other hand, we evaluated the
performance of different sets of observables for the readout layer. The results show that using correlations of the type $\braket{\sigma^a_i\sigma^b_j}$ with $a,b=x,y,z$ and $i\neq j$ can increase the total absolute capacity with respect to observables of the type $\braket{\sigma^a_i}$. This is the quantum advantage offered by exploiting the Hilbert space size of the quantum system. In particular, using virtual nodes and combinations of different sets of observables, saturation level of the total capacity is sustained. These results emphasize the importance of considering other sets of observables to boost the performance of quantum reservoir computers.

Quantum Reservoir Computing is a novel research field and there are several interesting open questions this work paves the way to address. The role of coherences, network topology and type of interactions are just a few examples. The main challenge that remains open is the assessment of measurements in the algorithm (back-action), beyond an ensemble picture. It will also be interesting to consider the extension of the formalism towards tasks with a quantum input. The former could lead to develop an algorithm which works in an online mode for temporal tasks for real experiments, while the latter would provide a full quantum machine learning approach, both for the processed input data and for the processing system.

 \vspace{0.5cm}

\noindent\textbf{Funding}
This study was financially supported by the Spanish State Research Agency, through the Severo Ochoa and Mar\'ia de Maeztu Program for Centers and Units of Excellence in R\&D (MDM-2017-0711), CSIC extension to QuaResC (PID2019-109094GB), and CSIC  Quantum Technologies PTI-001.
The work of RMP and MCS has been supported by MICINN, AEI, FEDER and the University of the Balearic Islands through a predoctoral fellowship (MDM-2017-0711-18-1), and a ``Ramon y Cajal” Fellowship (RYC-2015-18140), respectively. GLG acknowledges funding from the CAIB postdoctoral program.

\section*{Compliance with Ethical Standards}

\textbf{Conflict of Interest.} The authors declare that they have no conflict of interest.

\noindent\textbf{Ethical approval.} This article does not contain any studies with human participants or animals performed by any of the authors.

This is a pre-print of an article published in Cognitive Computation. The final authenticated version is available online at: \url{https://doi.org/10.1007/s12559-020-09772-y}

\section*{Appendix A}\label{app:Parameters}
\begin{figure}[htb]
\begin{center}
\includegraphics[width=\columnwidth]{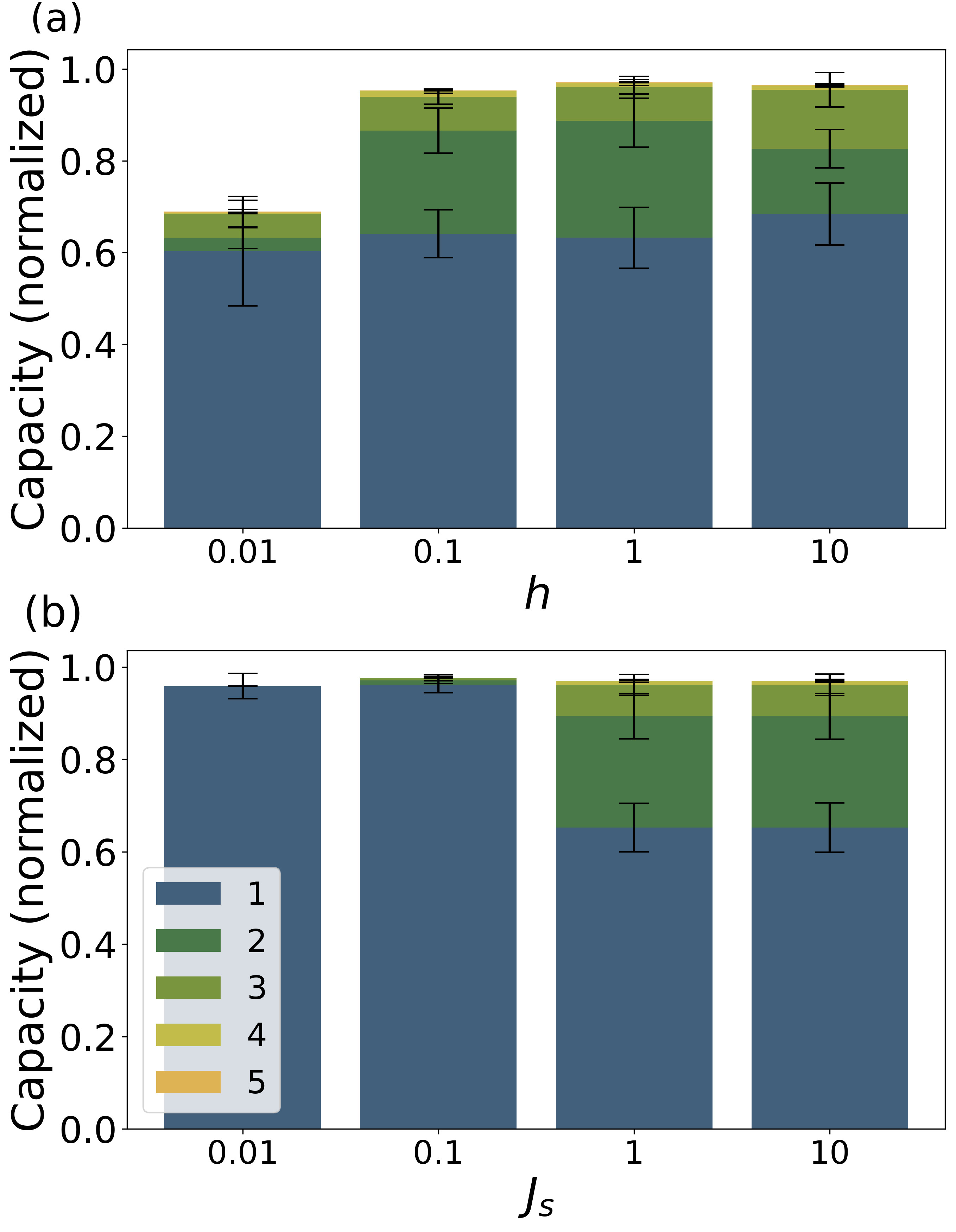}
\caption{IPC versus (a) natural frequency of the spins $h$ and (b) coupling strength $J_s$. The parameters in (a) are $\Delta t=10$, $J_s=1$ and $N=5$, while in (b) we have used $\Delta t=10$, $h=1$ and $N=5$. Notice that the normalization factor in this plot is $N$.} The error bars of the plots correspond to the standard deviation over 10 realizations.\label{Fig:8}
\end{center}
\end{figure}

Here, we motivate the choice of the parameters $h$ and $J_s$ of the model presented in Eq.~(\ref{Eq:H}).
Our choice is based on the numerical results shown in Fig.~\ref{Fig:8}, in which the IPC is computed for different values of $h$ and $J_s$.
Other relevant system parameters are set to $\Delta t=10$ and $N=5$, which have been our benchmark in the simulations according to the results presented in the main text. We have explored four orders of magnitude for both $h$ and $J_s$ to observe the evolution of the normalized capacity together with the distribution of the linear and nonlinear contributions to the IPC.
For small values of $h$, the total capacity does not saturate and the main contribution is the linear memory, being the only contribution for small values of $J_s$. For higher values of either $h$ or $J_s$, the profile of memory capacities changes towards a higher presence of nonlinear contributions, while keeping the bound of the total capacity.
Thus, the decision of choosing $h=1$ and $J_s=1$ as our benchmark (with $\Delta t=10$ and $N=5$) is based on the fact that it is a well established operational point, with a saturated total capacity, and a good presence of nonlinear contributions. 
\section*{Appendix B}
In this appendix, we explain in more detail how the bars of the IPC are computed.
Contributions to the IPC are usually shown according to the degree of the polynomial we want to reproduce.
For each degree, we need to sum up the contributions coming from different delays.
By delay we mean how far in the past we consider the influence of the input into the system. 
This influence is represented in Eq.~(\ref{Eq:Pol}) by taking the inputs $s_{k-i}$, where $i$ is the delay respect to present time $k$.

\begin{figure}[htb]
\begin{center}
\includegraphics[width=\columnwidth]{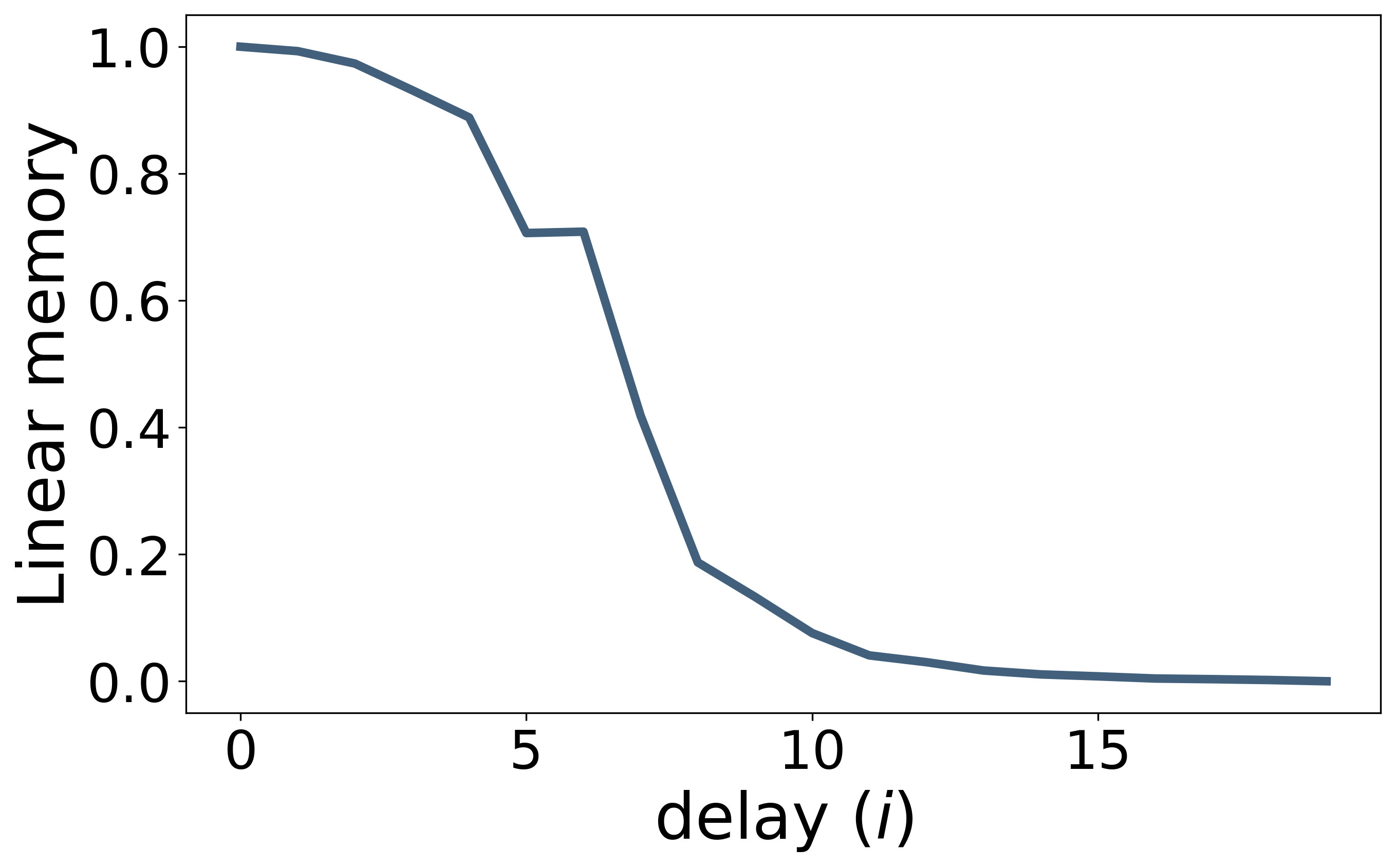}
\caption{Linear memory of the spin-based QRC for parameters $\Delta t=10$, $h=1$, $J_s=1$ and $N=5$ as a function of the delay in the input.}\label{Fig:9}
\end{center}
\end{figure}

In the main text, we have only shown the sum of the capacities over the delays. To deepen in our characterization we include here an illustration of the role of the delay for the reproduction of polynomials of degree 1, i.e. linear memory. 
The name of linear memory comes from the fact that we are computing the capacity of reproducing or ``remembering" targets of the form $\bar{y}_k=s_{k-i}$.
Fig.~\ref{Fig:9} represents the bare capacity of Eq.~(\ref{Eq:C}) with respect to delay $i$ for such a linear memory.  
The area under the curve of Fig.~\ref{Fig:9} is what we have plotted as IPC of degree $d=1$ in the bar's plots across this work. 
In some cases, e.g. Fig.~\ref{Fig:8} (b) ($J_s=0.01$), the influence of the input extends to delays longer than 100 past inputs and care needs to be taken to not disregard small non-vanishing contributions in the computation of the IPC.
It is a straightforward procedure to represent linear memory, but nonlinear contributions are harder to untangle visually. Representing second and third-order contributions is still possible with 2D and 3D heatmaps of the capacities respect to the multiple delays. However, such visual representations of the memory as a function of the delay find a limit when we go to nonlinearities of degree $d\geq 4$. Therefore, although we are aware that the summation over delay contributions is somehow hiding the information regarding the distribution of the memory for different delays, it provides a compact and readable representation. 

\bibliography{arxiv}

\begin{thebibliography}{10}

\bibitem{goodfellow2016deep}
Goodfellow I, Bengio Y, Courville A.
\newblock Deep learning.
\newblock MIT press; 2016.

\bibitem{krizhevsky2012imagenet}
Krizhevsky A, Sutskever I, Hinton GE.
\newblock Imagenet classification with deep convolutional neural networks.
\newblock In: Advances in Neural Information Processing Systems; 2012. p.
  1097--1105.

\bibitem{carleo2019machine}
Carleo G, Cirac I, Cranmer K, Daudet L, Schuld M, Tishby N, et~al.
\newblock Machine learning and the physical sciences.
\newblock Reviews of Modern Physics. 2019;91(4):045002.

\bibitem{hinton2018deep}
Hinton G.
\newblock Deep learning—a technology with the potential to transform health
  care.
\newblock Jama. 2018;320(11):1101--1102.

\bibitem{young2018recent}
Young T, Hazarika D, Poria S, Cambria E.
\newblock Recent trends in deep learning based natural language processing.
\newblock IEEE Computational Intelligence magazine. 2018;13(3):55--75.

\bibitem{triefenbach2014large}
Triefenbach F, Demuynck K, Martens JP.
\newblock Large vocabulary continuous speech recognition with reservoir-based
  acoustic models.
\newblock IEEE Signal Processing Letters. 2014;21(3):311--315.

\bibitem{pathak2018model}
Pathak J, Hunt B, Girvan M, Lu Z, Ott E.
\newblock Model-free prediction of large spatiotemporally chaotic systems from
  data: A reservoir computing approach.
\newblock Physical Review Letters. 2018;120(2):024102.

\bibitem{antonik2016online}
Antonik P, Duport F, Hermans M, Smerieri A, Haelterman M, Massar S.
\newblock Online training of an opto-electronic reservoir computer applied to
  real-time channel equalization.
\newblock IEEE Transactions on Neural Networks and Learning Systems.
  2016;28(11):2686--2698.

\bibitem{makridakis2018m4}
Makridakis S, Spiliotis E, Assimakopoulos V.
\newblock The M4 Competition: Results, findings, conclusion and way forward.
\newblock International Journal of Forecasting. 2018;34(4):802--808.

\bibitem{maass2002real}
Maass W, Natschl{\"a}ger T, Markram H.
\newblock Real-time computing without stable states: A new framework for neural
  computation based on perturbations.
\newblock Neural Computation. 2002;14(11):2531--2560.

\bibitem{lukovsevivcius2009reservoir}
Luko{\v{s}}evi{\v{c}}ius M, Jaeger H.
\newblock Reservoir computing approaches to recurrent neural network training.
\newblock Computer Science Review. 2009;3(3):127--149.

\bibitem{jaeger2001echo}
Jaeger H.
\newblock The “echo state” approach to analysing and training recurrent
  neural networks-with an erratum note.
\newblock Bonn, Germany: German National Research Center for Information
  Technology GMD Technical Report. 2001;148(34):13.

\bibitem{verstraeten2007experimental}
Verstraeten D, Schrauwen B, d’Haene M, Stroobandt D.
\newblock An experimental unification of reservoir computing methods.
\newblock Neural Networks. 2007;20(3):391--403.

\bibitem{tanaka2019recent}
Tanaka G, Yamane T, H{\'e}roux JB, Nakane R, Kanazawa N, Takeda S, et~al.
\newblock Recent advances in physical reservoir computing: A review.
\newblock Neural Networks. 2019;.

\bibitem{van2017advances}
Van~der Sande G, Brunner D, Soriano MC.
\newblock Advances in photonic reservoir computing.
\newblock Nanophotonics. 2017;6(3):561--576.

\bibitem{appeltant2011information}
Appeltant L, Soriano MC, Van~der Sande G, Danckaert J, Massar S, Dambre J,
  et~al.
\newblock Information processing using a single dynamical node as complex
  system.
\newblock Nature Communications. 2011;2(1):1--6.

\bibitem{torrejon2017neuromorphic}
Torrejon J, Riou M, Araujo FA, Tsunegi S, Khalsa G, Querlioz D, et~al.
\newblock Neuromorphic computing with nanoscale spintronic oscillators.
\newblock Nature. 2017;547(7664):428.

\bibitem{fujii2017harnessing}
Fujii K, Nakajima K.
\newblock Harnessing disordered-ensemble quantum dynamics for machine learning.
\newblock Physical Review Applied. 2017;8(2):024030.

\bibitem{nakajima2019boosting}
Nakajima K, Fujii K, Negoro M, Mitarai K, Kitagawa M.
\newblock Boosting computational power through spatial multiplexing in quantum
  reservoir computing.
\newblock Physical Review Applied. 2019;11(3):034021.

\bibitem{chen2019learning}
Chen J, Nurdin HI.
\newblock Learning nonlinear input--output maps with dissipative quantum
  systems.
\newblock Quantum Information Processing. 2019;18(7):198.

\bibitem{ghosh2019quantum}
Ghosh S, Opala A, Matuszewski M, Paterek T, Liew TC.
\newblock Quantum reservoir processing.
\newblock npj Quantum Information. 2019;5(1):1--6.

\bibitem{markovic2020quantum}
Markovi{\'c} D, Grollier J.
\newblock Quantum neuromorphic computing.
\newblock arXiv preprint arXiv:200615111. 2020;.

\bibitem{nielsen2010quantum}
Nielsen MA, Chuang I. Quantum computation and quantum information.
\newblock Cambridge University Press; 2010.

\bibitem{ladd2010quantum}
Ladd TD, Jelezko F, Laflamme R, Nakamura Y, Monroe C, O’Brien JL.
\newblock Quantum computers.
\newblock Nature. 2010;464(7285):45--53.

\bibitem{acin2018quantum}
Ac{\'\i}n A, Bloch I, Buhrman H, Calarco T, Eichler C, Eisert J, et~al.
\newblock The quantum technologies roadmap: a European community view.
\newblock New Journal of Physics. 2018;20(8):080201.

\bibitem{dambre2012information}
Dambre J, Verstraeten D, Schrauwen B, Massar S.
\newblock Information processing capacity of dynamical systems.
\newblock Scientific Reports. 2012;2:514.

\bibitem{jaeger2002short}
Jaeger H.
\newblock Short term memory in echo state networks. GMD-Report 152.
\newblock In: GMD-German National Research Institute for Computer Science
  (2002). Citeseer; 2002. .

\bibitem{brunner2013parallel}
Brunner D, Soriano MC, Mirasso CR, Fischer I.
\newblock Parallel photonic information processing at gigabyte per second data
  rates using transient states.
\newblock Nature Communications. 2013;4:1364.

\bibitem{larger2017high}
Larger L, Bayl{\'o}n-Fuentes A, Martinenghi R, Udaltsov VS, Chembo YK, Jacquot
  M.
\newblock High-speed photonic reservoir computing using a time-delay-based
  architecture: Million words per second classification.
\newblock Physical Review X. 2017;7(1):011015.

\bibitem{jaeger2004harnessing}
Jaeger H, Haas H.
\newblock Harnessing nonlinearity: Predicting chaotic systems and saving energy
  in wireless communication.
\newblock Science. 2004;304(5667):78--80.

\bibitem{soriano2014delay}
Soriano MC, Ort{\'\i}n S, Keuninckx L, Appeltant L, Danckaert J, Pesquera L,
  et~al.
\newblock Delay-based reservoir computing: noise effects in a combined analog
  and digital implementation.
\newblock IEEE Transactions on Neural Networks and Learning Systems.
  2014;26(2):388--393.

\bibitem{grigoryeva2018echo}
Grigoryeva L, Ortega JP.
\newblock Echo state networks are universal.
\newblock Neural Networks. 2018;108:495--508.

\bibitem{negoro2018machine}
Negoro M, Mitarai K, Fujii K, Nakajima K, Kitagawa M.
\newblock Machine learning with controllable quantum dynamics of a nuclear spin
  ensemble in a solid.
\newblock arXiv preprint arXiv:180610910. 2018;.

\bibitem{chen2020temporal}
Chen J, Nurdin HI, Yamamoto N.
\newblock Temporal information processing on noisy quantum computers.
\newblock arXiv preprint arXiv:200109498. 2020;.

\bibitem{nokkala2020gaussian}
Nokkala J, Mart{\'\i}nez-Pe{\~n}a R, Giorgi GL, Parigi V, Soriano MC, Zambrini
  R.
\newblock Gaussian states provide universal and versatile quantum reservoir
  computing.
\newblock arXiv preprint arXiv:200604821. 2020;.

\bibitem{keuninckx2017real}
Keuninckx L, Danckaert J, Van~der Sande G.
\newblock Real-time audio processing with a cascade of discrete-time delay
  line-based reservoir computers.
\newblock Cognitive Computation. 2017;9(3):315--326.

\bibitem{gallicchio2017echo}
Gallicchio C, Micheli A.
\newblock Echo state property of deep reservoir computing networks.
\newblock Cognitive Computation. 2017;9(3):337--350.

\bibitem{Rodrigo}
Mart{\'\i}nez-Pe{\~n}a R, Giorgi GL, Nokkala J, Zambrini R, Soriano MC.
  Dynamical phase transitions in quantum reservoir computing.
\newblock in preparation;.

\end{thebibliography}
\bibliographystyle{vancouver}
\end{document}